\documentclass[runningheads]{llncs}

\usepackage{vdm-rg}
\usepackage{xfrac}
\leftRecord
\leftCases
\usepackage{textcomp}

\usepackage{xcolor}

\usepackage{hyperref}
\usepackage{enumitem}

\newcommand{\defeq}{\stackrel{def}{=}}

\usepackage{csquotes}

\begin{document}

%\title{Using R/G Conditions to model Mixed Criticality notions of Robustness and Resilience}

\title{A Rely-Guarantee Specification of Mixed-Criticality Scheduling}
%\title{Reasoning about the Relationship Between the Scheduler and Mixed-Criticality Jobs}

%\author{Cliff B Jones\inst{1} \and Alan Burns\inst{2}}

\author{Cliff B Jones\inst{1} \and
Alan Burns\inst{2}}
\institute{School of Computing, Newcastle University, UK \and
Department of Computer Science, University of York, UK}

%Princeton University, Princeton NJ 08544, USA \and
%Springer Heidelberg, Tiergartenstr. 17, 69121 Heidelberg, Germany
   
\maketitle

\noindent 
\makebox[\linewidth]{\today} \\

%\noindent
%\makebox[\linewidth]{\small 2020-12-01}

\begin{abstract}
The application considered here is mixed-criticality scheduling.
The core formal approaches used are Rely-Guarantee conditions and the Timeband framework;
these are applied to give a layered description of job scheduling
which covers resilience to jobs overrunning their expected execution time.
A novel formal modelling idea is proposed to handle the relationship between actual time and its approximation in hardware clocks.
\end{abstract}

\begin{minipage}{0.7\textwidth}
{\bf Note}\\
This paper is an updated version of the one
 written for {\em Mathematical Foundations of Software Engineering: Essays in Honor of Tom Maibaum on the Occasion of his Retirement} 
 (eds. 
 Nazareno Aguirre,
Valentin Cassano,
Pablo Castro,
Ramiro Demasi) 
College Publications Tributes Series, 2021.
\end{minipage}

\vspace{4ex}

\section{Introduction}\label{intro}

The objective of the research described in this  paper is to develop a framework based on time bands and rely-guarantee conditions
for formally specifying and reasoning about properties of mixed-criticality scheduling (MCS);
the key correctness issues revolve around timing.
%The background ideas are summarised in Section~\ref{S-approaches} after the application area is outlined in Section~\ref{S-MCS}.
%repeated below

%\cnote{The following sections \ldots}
Section~\ref{S-MCS} outlines the application area
while Section~\ref{S-approaches} briefly sketches published ideas on the timeband framework and
rely-guarantee thinking together with
the application of both of these formalisms to the specification of cyber physical systems.
Section~\ref{S-time} tackles the thorny issue of the passage of time.
%(see the quotation at the head of this paper):
%coming up with a satisfactory way to relate real physical time to values manipulated by programs 
%held up writing this paper for some \ldots\ time!
These ideas are brought together in Section~\ref{S-jobs} to write layered descriptions of MCS
covering both optimistic and resilient modes.
The customary summary and statement of future work are given in Section~\ref{S-con}.

%\alannote{I think the intro needs a bit more on R/G - as this is central to the paper, or
%a section on R/G before the following one on MCS.}
%
%\cnote{Alan: I've added a footnote pointing to the descriptions;
%I did consider reversing the order of Sections~\ref{S-approaches}/\ref{S-MCS} but decided against}

%%%%%%%%%%%%%%%%%%%%%%%%%%%%
\section{Mixed-criticality scheduling}
\label{S-MCS}

The implementation of any real-time system requires a run-time scheduler that will follow
the rules that define the required behaviour of the scheduling approach that has been
chosen to deliver the temporal properties of the application. A human scheduling specialist
will choose the basic approach, for example a fixed priority scheme with priorities assigned
via the deadline monotonic algorithm, or the Earliest Deadline First protocol. Appropriate scheduling
analysis will then be applied to a specification of the application to determine whether all deadlines can
be met at run-time by the chosen scheduling approach.

Typical characteristics of an application are the number of jobs or tasks involved, the worst-case
execution times of these entries, their deadlines and possible constraints on their arrival patterns.
The dynamic run-time scheduler will rely on the validity of these parameters
and will guarantee to manage the order of execution in accordance with the rules of the chosen scheduling protocol.

An increasingly important trend in the design of real-time systems is the
integration of components with different levels of criticality on a common hardware
platform. Criticality is a designation of the level of assurance
against failure needed for a system component. A mixed-criticality system is
one that has two or more distinct levels (for example safety critical, mission
critical and low-criticality). Perhaps up to five levels may be identified
(see, for example, the IEC 61508, DO-178B and DO-178C, DO-254 and ISO 26262 standards).
Typical names for the levels are ASILs (Automotive Safety and Integrity Levels), DALs (Design Assurance Levels
or Development Assurance Levels)
and SILs (Safety Integrity Levels).

A key aspect of MCS is that system parameters, such as estimates of the worst-case execution time (WCET) of a job,
become dependent on the criticality level of the job. So the same code will have higher WCET estimates
for jobs defined to be safety-critical ---
this is necessary as a higher level of
assurance is required than it would be if it is just considered to be mission critical or indeed non-critical. 
Criticality also has a role when the
system becomes overloaded; the jobs of a lower level of criticality may have to
be abandoned to provide resources to the safety-critical (high integrity) jobs.

A mixed criticality (MC-) scheduler is one that manages a set of mixed criticality jobs or tasks
(the current paper addresses only jobs --- 
extensions to address tasks are mentioned in Section~\ref{S-impl}) 
so that all deadlines are met if all jobs execute for no more than a lower
bound on their execution time. In addition, the MC-scheduler must ensure that all high
criticality jobs meet their deadlines if any job executes for more than its lower bound
(but less than a conservative upper bound defined for each job).

%\cnote{Alan: do you agree these paragraphs would sit better in \S\ref{S-MCS}?}

The following paragraphs relate this objective to the wider topic of Cyber Physical Systems (CPS).

{\bf Correctness}  in safety-critical CPS can be considered from two
perspectives: (i) (pre-run-time) verification, and (ii) survivability.
Pre-run-time {\em verification\/} of a safety-critical system is the
process of ensuring, prior to deployment, that the run-time
behaviour of the system will be consistent with expectations.
Verification assumptions are made regarding
the kinds of circumstances that will be encountered by the
system during run-time and guarantees are used to specify the required runtime
behaviour of the system (provided
that the assumptions hold).

In contrast, {\bf survivability\/} addresses
expectations of system behaviour in the event that the assumptions
fail to hold in full (in which case a fault or error is said
to have occurred during run-time). Survivability may further
be considered to comprise two notions: robustness and
resilience~\cite{8352522}. Informally, the robustness of a system is a measure
of the severity and number of faults it can tolerate without compromising
the quality of service it offers while resilience refers
to the degree of fault for which it can provide degraded yet
acceptable quality of service.

%\alannote{appropriate scheduling is a mixture of (static) decisions made by a human scheduling expert
%and a dynamic scheduling algorithm  selected at the time the schedule is considered\\
%static scheduling relies on information about (probable) arrival patterns of jobs\\
%a typical run-time algorithm  is ``earliest deadline first''}

It must be emphasised that the internal details of MC-scheduler, and the theory used to
define the associated schedulability analysis (for example the EDF-VD or AMC protocols) are not the emphasis in this paper --- 
there is plenty of prior research on that area~\cite{burns:survey:2017}. 
Nor is there an explanation of how previously-proposed
MC-scheduling algorithms can be shown to satisfy particular sets of rely-guarantee (R/G) specifications --- 
that is (important) future work. 
This paper only seeks to provide a clear and intuitive specification of the components and thus motivate the formalism. 
The history of formal methods (such as Hoare Logic) prompts the belief
that methods can be developed for showing that specific
MC-scheduling algorithms satisfy (or fail to so do) particular R/G specifications.

%%%%%%%%%%%%%%%%%%%%%%%%%%%%
\section{Background approaches}
\label{S-approaches}

This section describes previously published ideas on which the developments in this paper are based.

%%%%%%%%%%%%%%%
\subsection{Rely-guarantee reasoning for concurrency}
\label{S-RG}

Pre and post conditions are used to document the intended behaviour of sequential programs.
Such specifications can be said to document the ``Why'' rather than the ``How'' of a component.
Furthermore,
developments of Tony Hoare's ``axiomatic approach''~\cite{Hoare69a,Hoare71a}
provide ways of evolving verified implementations in a top-down style.
(Even if the development is not actually undertaken in this way,
such a structure provides understandable documentation.)
Key to layering such a description is a property that is often referred to as ``compositionality'':
the specification of a component describes all that need be achieved by its implementation;
such specifications insulate a component from considerations about its environment
and facilitate the verification of one design step before proceeding to further stages of development.

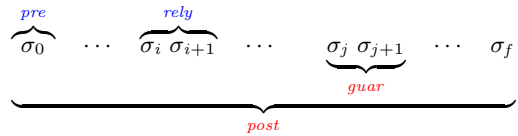
\begin{figure}
\begin{center}
$
\underbrace{\overbrace{\sigma\sb{0}}^{\color{blue}pre\color{black}} \hspace{1em} \cdots \hspace{1em} \overbrace{\sigma\sb{i} \; \sigma\sb{i+1}}^{\color{blue}rely\color{black}} \hspace{1em} \cdots \hspace{2em} \underbrace{\sigma\sb{j} \; \sigma\sb{j+1}}\sb{\color{red}guar\color{black}} \hspace{1em} \cdots \hspace{1em} \sigma\sb{f}}\sb{\color{red}post\color{black}}
$
\end{center}
%P/R as assumptions for designer\\
%G/Q as commitments on code
{\color{blue} $pre/rely$ are assumptions the developer can make}

{\color{red} $guar/post$ are commitments that the code must achieve}

\caption{Picturing the parts of a Rely-Guarantee specification}
\label{F-RG}

\end{figure}

Finding compositional development methods for concurrent software proved challenging with many initial methods
(e.g.~\cite{AshcroftManna71,Ashcroft75,Owicki75,OwickiGries76})
needing to reason jointly about the combination of one thread with its sibling threads and/or environment.
This frustrates the ability to achieve top-down design.
The Rely-Guarantee approach~\cite{Jones81d,Jones83a,Jones83b} 
offers compositional specifications for a class of shared-variable concurrent programs.
Just as pre conditions record assumptions about the context in which a component can be deployed,
a rely condition indicates what interference a component must tolerate.
Thus pre and rely conditions are information to the developer and warnings to anyone who wishes to deploy the specified component.
A similar comparison can be made between the two conditions that record properties that the developed code must satisfy:
post conditions describe the relation required of starting and finishing states%
\footnote{Pre and post conditions are as in~\cite{Hoare69a} except that in VDM~\cite{Jones80a}
post conditions are relations over initial and final states.}
whereas guarantee conditions
indicate an upper bound on the impact that steps of the component can have on its environment.
A picture of these components of a specification is given in Figure~\ref{F-RG}.

In~\cite{Hoare69a}, Hoare offered proof rules that justify development using the main sequential programming constructs.
It is not surprising that the proof rules which justify steps of development employing parallelism are more complicated:
they essentially need to show the compatibility of the rely and guarantee conditions.%
\footnote{A significant reworking~\cite{HayesJonesColvin14TR,Jones-FACJexSEFM-13,HayesJones18}
presents the original rely-guarantee ideas in a more algebraic style.}
The current paper does not go into these details because description rather than proof is the objective here.

One observation that does carry over from development methods for sequential programs
to many applications of the Rely-Guarantee approach is the importance of data abstraction and reification~\cite{Jones06a}.
Although predicates over states provide an element of procedural abstraction,
specifications of significant systems can only be made brief and perspicuous by using abstract data types that match the problem.
Subsequent development steps must show that representations of the abstractions preserve the properties of the specification.
Development and justification of more concrete representations is variously referred to as ``refinement'' or ``reification''.

\subsection{Time bands}

The motivation for the timeband framework comes from a number of observations
about complex time-sensitive systems.
Of relevance to this paper are the following:
\begin{itemize}
  \item systems can be best understood by distinguishing different granularities (of
  time), i.e. there are different abstract views of the dynamics of
  the system
  \item it is useful to view certain actions (events) as atomic and ``instantaneous'' in one time band,
  while allowing them to have internal state and behaviour that takes time at a more
  detailed level of description
  \item the durations of certain actions are important, but the
  measuring of time must not be made artificially precise and must allow for tolerance 
  in the temporal domain.
  \end{itemize}

\noindent
Key references include~\cite{BurnsHayesJones-19,BurnsHayes09,R:Wei:2012,R:Woodcock:2010,R:Baxter:2007,timebandbook:2006}.

The central notion in the framework is that of a time band that is
defined by its \emph{granularity}, $G$,
(e.g.~1 millisecond)
and its \emph{precision}, $\rho$,
(e.g.~ 5 microseconds).
Granularity defines the unit of time of the band; precision bounds the
maximum duration of an event that is deemed to be instantaneous in its
band.

Whilst it is the case that system descriptions can be given on a single time axis,
inevitably, this has to be a fine granularity and it becomes difficult to ``see the wood for the trees.''
It is much clearer if the behaviour of a system is given in terms of a finite set of ordered \emph{bands}.
System activities are described in some band
$B$ if they engage in significant events at the time scale represented
by $B$, i.e.\ they have dynamics that give rise to changes that are
observable or meaningful in the granularity of band $B$.

A complete system specification must address all dynamic behaviours. At the lowest level,
circuits (e.g.\ gates) have propagation delays measured in nanoseconds or even picoseconds; 
at intermediate levels,
tasks/threads have rates and deadlines that are usually expressed in tens of milliseconds;
at yet-higher levels, missions can change every hour; and maintenance may need to be
undertaken every month.

Understanding the behaviour of circuits allows the worst-case
execution time of tasks to be predicted. At a higher band
this allows deadlines to be checked and
the schedulability of whole missions to be verified.

In this paper,
behaviour of the application jobs and the scheduler is placed in
a band that reflects the deadlines of the jobs; this might be a band with a granularity of
one millisecond, or a finer granularity if that is required. The precision in this band will
be sufficiently short so that the duration of certain actions, for example a context switch, can be ignored.

Precision is employed, in Section~\ref{S-ClockValue},  to constrain the difference between external `real' time
and any interval interpretation of time as delivered by a hardware clock.

%%%%%%%%%%%%%%%
\subsection{Specifying resilient CPS}

As indicated in Section~\ref{S-RG},
the Rely-Guarantee approach was originally conceived for developments where a specified system was to be decomposed into concurrent processes.
The general idea has however been shown to be applicable to contexts where a component is being specified 
which will execute in an environment that evolves in parallel with the specified component.
%\cnote{Alan: I'd like to ad a sentence/paragraph on ``deriving the specification of ``control programs'''' ---
%this would link to the references to ``control'' below}
Examples in~\cite{HayesJacksonJones03,JonesHJ07}
indicate how the specification of the control component can be derived from a specification of an overall cyber-physical system
by recording assumptions (rely conditions) about the physical components.
These early papers suffered from the fact that time was treated on a single
(i.e.~the finest)
band;
in~\cite{BurnsHayesJones-19},
the timeband approach was used to make the rely and guarantee conditions more intelligible.
This indicated that combining the timeband and rely-guarantee approaches can be used to specify CPS.
A particular issue with CPS is that they need to be resilient in the sense that 
they are likely to have layers of required behaviour:

\begin{itemize}

\item optimistic (or optimal) behaviour is required when everything is performing in accord with the strongest rely conditions ---
the control system is required to meet its strongest guarantee condition;

\item when some rely conditions are not satisfied,
something in the environment is not behaving in an optimal way
(this can be caused by a timing problem) ---
a weaker rely condition can describe a less desirable environment assumption under which the control system
can only achieve a weaker guarantee condition;

\item such layering of rely and guarantee conditions can be repeated over as many levels as required.

\end{itemize}

\noindent
Such nested conditions 
(combined with time bands)
are illustrated in~\cite[\S4]{BurnsHayesJones-19}.
Another idea that appears to be useful in specifying CPS is specifying ``may/must'' constraints:
\cite[\S4.3]{BurnsHayesJones-19} contains an example where a short period of aberrant behaviour can be flagged 
but the control system is required to report a longer period of misbehaviour.
This pattern of specification appears to be useful in a number of situations
(again often linked to time).

%%%%%%%%%%%%%%%%%%%%%%%%%%%%
\section{Handling time}
\label{S-time}

Specifying the sort of scheduling problem described in Section~\ref{S-MCS}
presents additional challenges not faced in,
for example,~\cite{BurnsHayesJones-19}:

\begin{itemize}

\item internal machine clocks must be linked to the passage of actual time

\item state variables are needed that record the amount of time used by a job.

\end{itemize}

\noindent
What follows is a novel approach to these two problems.

% linking the passage of time with the values recorded in computer clocks.
%The simple scheduling task considered in the body of the paper concerns a finite number of ``one shot'' jobs;

%%%%%%%%%%%%%%
\subsection{Relating $ClockValue$ to $Time$}
\label{S-ClockValue}

The first step is to distinguish an abstract notion of $Time$ from what clocks record in a computer
(a clock will contain a  $ClockValue$).
Consider a collection of $State$s indexed by the abstract notion of $Time$:

\type{\Sigma}{Time \to State}

\noindent
is a function;
$Time$ should be ``dense'' like the real numbers 
(so $\Sigma$ cannot be modelled as a list)
--- but fortunately it transpires that little need be said about $Time$ 
because specifications of operations 
(e.g.~the scheduler and the jobs that it controls) 
are written with respect to the $t$ component of $State$:

\begin{record}{State}
t: ClockValue\\
\cdots
\end{record}

\noindent
For a given $\sigma: \Sigma$ at time $\alpha$
(because the identifier $t$ is used for a component of $State$,  $\alpha \in Time$ is used here),
its $ClockValue$ ($\sigma(\alpha).t$) can differ from $\alpha$.
This is an issue for a fine time band
and time bands need to be chosen such that the allowable difference is within the precision of the coarser band
so that it can essentially be assumed that the $ClockValue$ in the machine is always sufficiently close to $Time$.%
\footnote{At the finer time band,
issues such as clock drift could also be formalised.
In a distributed system,
there would be a local $State$ for each processor and their clocks could also differ within the appropriate precision.}

Assume that the precision  of the band is $\rho$,
equality $=\sb{\rho}$ is with respect to that precision.
Then the relationship between $t$ and $\alpha$ is defined by the following predicate:

\begin{formula}
P\sb{t}(\sigma) \defeq 
  \forall{\alpha \in Time}{\sigma(\alpha). t = \sb{\rho} \alpha}
\end{formula}

%%%%%%%%%%%%%%
\subsection{Tracking execution time}
\label{S-exec}

The set $State$ can for the most part be viewed as in any model-oriented specification in say VDM or Event-B.
Specified software operations cause changes from one state to another value in $State$.

\begin{record}{State}
t: ClockValue\\
active: \mapof{I}{Job}
\end{record}

\begin{record}{Job}
e: Duration\\
run: \Bool\\
info: JobInfo
\end{record}

\noindent
Information about any $Job$ that has started and not finished is stored in the $active$ map.
If the job is running at $ClockValue$ (time) $t$ then the Boolean flag $run$ is
true (otherwise it is false).
The Scheduler controls when a job is running by flipping the $run$ flag.%
\footnote{To model multi-processors, this can be the case with more than one $Job$.}
($JobInfo$ is discussed below.)

The predicate $P\sb{t}$ above defines how $t$ values relate to $Time$ but
the specifications of jobs and Scheduler also need to refer to the execution time of a job:
the $e$ field of a $Job$ records the summation of its execution time up to $ClockValue$ $t$.

Neither the scheduler nor any job can change $t$ or $e$
which are instead linked to the autonomous 
(i.e.~not under the control of software)
progress of time:
any job that is running over a period of time from $\alpha\sb{1}$ to $\alpha\sb{2}$ will have its $e$ field 
advanced by as much as the difference in the $Time$%
\footnote{The difference between two values of type $Time$ is a $Duration$ 
as is the difference between two values of type $Duration$.
$Duration$s are non-negative.}
i.e.~$e\sb{2} \minus e\sb{1}$ must be equal, within the precision of the time band, to $\alpha\sb{2} \minus \alpha\sb{1}$;
furthermore, when job $i$ is not running,
its $e$ field remains unchanged.
Thus the link between $Time$ and $e$ is defined by a predicate over $\Sigma$:

\begin{formula}
P\sb{e}(\sigma) \defeq \T2
  \forall*{\alpha\sb{1}, \alpha\sb{2} \in Time, i \in Index}{
%  	\begin{formbox}
  	\begin{formbox}
	\hspace{3ex}((\forall{\alpha | \alpha\sb{1} \leq \alpha \leq \alpha\sb{2}}{(\sigma(\alpha).active)(i).run}) \Implies \T4
 			\sigma(\alpha\sb{2}).active(i).e \minus \sigma(\alpha\sb{1}).active(i).e =\sb{\rho} \alpha\sb{2} \minus \alpha\sb{1}) \And
	\end{formbox}\\
  	\begin{formbox}
	\hspace{3ex}((\forall{\alpha | \alpha\sb{1} \leq \alpha \leq \alpha\sb{2}}{\Not(\sigma(\alpha).active)(i).run}) \Implies \T4
 			\sigma(\alpha\sb{2}).active(i).e = \sigma(\alpha\sb{1}).active(i).e) 
	\end{formbox}
%	\end{formbox}
	}
\end{formula}

Although $Time$ is dense and progresses outside the influence of the software
($Scheduler$ or $Job$s),
it is precisely that software that brings about discrete changes to $State$.
The specifications of the software components are written with respect to $State$ and thus relate to $t: ClockValue$
but that $t$ component is changed by the progress of $Time$.
Essentially,
real time is about the $\Sigma$ function
whereas programs actually bring about discrete changes in $State$s
(except, of course, neither $Scheduler$ nor $Job$s can write to $t$ or $e$ whose values are constrained 
by $P\sb{t}/P\sb{e}$ above).

%\cnote{HAVE I ANSWERED:\\
%Alan's innocent question about linking $run$ with $t$ might actually require a lot of explanation!\\
%%Between two $State$s changed by the software, there will be an infinity of $State$s with updated $t$ and $e$ per $Job$!
%%Those differences are mediated by $P\sb{e}/P\sb{t}$.\\
%%But it remains the case that fields like $run$ are only changed by software.
%}
%\anote{I've added a little to start of section 4.2 - OK?}

%\cnote{There are really 3 (sorts of) processes: $scheduler || Jobs || Time$}

Although the focus in Section~\ref{S-jobs} is on specifications for the scheduler and the jobs to which it is allocating time,
the progress of $Time$ is really a third concurrent process.
What could be thought of as a guarantee condition of this enigmatic process 
(the conjunction of $P\sb{e}$ and $P\sb{t}$)
can be used as an assumption in any reasoning about the components 
($Scheduler$ and $Job$s)
that are to be programmed.

%%%%%%%%%%%%%%%%%%%%%%%%%%%%
\section{Job-Based Scheduling}
\label{S-jobs}

%\todo{In outline:\\
%static vs. dynamic scheduling\\
%R/G of whole scheduling process would include assumptions about arrival of Jobs\\
%(optimistic) R/G for dynamic scheduler + G of Jobs\\
%(resilient) R/G accepts Jobs might overrun + higher estimate for hi-crit Jobs + G only those get done\\
%note that if resilient R on hi-crit Jobs is not satisfied, no G\\
%Note that Tasks are a larger structure than Jobs -- but Tasks not considered here}
As outlined in Section~\ref{S-MCS},
scheduling work can be divided into two parts.
A human scheduler performs ``schedulability analysis'' 
which considers the likely arrival pattern --and estimates of the {\em worst case execution time}-- of jobs%
\footnote{In practical scheduling applications, 
there is likely to be a {\em task level} above jobs 
but this is not addressed in the current paper.}
%It is addressed in the companion~\cite{BurnsJones-RTSS-21}.}
and chooses a scheduling algorithm for the run-time scheduling software (which comprises the second part).
The aim of the Scheduler software component is to ensure that jobs complete execution by their respective deadlines.
The rely conditions of the combined scheduling activities would detail the inputs to the static part of scheduling.
In this section,
the focus is on specifying the run-time scheduling software using rely and guarantee conditions.

In safety-critical situations,
resilience is crucial and minor deviation from the estimated run times must not be allowed to 
cause highly-critical (hi-crit) jobs to miss their deadlines.
The categorisation of jobs as hi/lo-crit is part of the off-line schedulability analysis.%
\footnote{In reality, there could be many levels of criticality but the approach can be illustrated with just two.}
Here,
nested rely and guarantee conditions are used to specify an optimistic mode in which all jobs can meet their deadlines
and a fault-tolerant mode in which only hi-crit jobs are guaranteed to meet their deadlines.

The overall run-time system can be considered to consist of:

\begin{formula}
Passage-of-Time || SCHEDULER || \; ||\sb{i} JOB(i) 
\end{formula}

\noindent
But it is important to understand the ``frames'' of these components.
Starting with the enigmatic $Passage-of-Time$
this is the only component that relates to $\Sigma$ and it guarantees both $P\sb{t}$ and $P\sb{e}$
(recall that $P\sb{t}$ links $\alpha \in Time$ with $\sigma(\alpha).t$).

The $SCHEDULER$ has a frame  of $State$ because it has purview over the whole collection of $JOB$s.
An individual $JOB(i)$ only has access to its own $Job$ information.

The run-time Scheduler 
%(denoted by the symbol $S$)
manages the execution of
$Job$s
(indexed by $i \in I$ --- the set $I$ is an arbitrary index set).
As indicated in Section~\ref{S-exec},
$active$ contains information about those jobs that have started and not yet finished.

As defined above, the local state for each job is as follows:

\begin{record}{Job}
e: Duration\\
run: \Bool\\
info: JobInfo
\end{record}

\noindent
As mentioned in Section~\ref{S-ClockValue},
the time ($e$) that a job has been executing cannot be changed directly by the scheduler and is updated in accordance with $P\sb{e}$.

Static information about a $Job$ is contained in:

\begin{record}{JobInfo}
d: ClockValue\\
C: Duration\\
crit: \const{Lo} | Hi
\end{record}

\noindent
The deadline by which a job should complete ($d$) is set when a job starts (see below) ---
it is shown to be of type $ClockValue$ because,
although deadlines relate to the external world,
software can only access the internal timers 
(thus $t$) ---
but $P\sb{t}$ ensures that software can use $t: ClockValue$ as an acceptable surrogate.
The $C$ field contains the expected maximum execution time of the job.
Finally $crit$icality can either be a simple flag ($\const{Lo}$) or, 
for high criticality jobs, 
can store an extra allocation that can be used if they overrun:
 
\begin{record}{Hi}
X: Duration
\end{record}

%%%%%%%%%%%%%%
\subsection{Starting and ending $Job$s}

The ``life cycle'' of a $JOB$ is:

\begin{itemize}

\item A $JOB$ appears (not controlled by the $SCHEDULER$) when it enters $active$;
this happens at $\alpha \in Time$; this time is registered internally
as $ClockValue, t$ (with $t =\sb{\rho} \alpha$).
The execution time ($e$) at the start of a job is set to zero;
and its absolute deadline $d$ is set to $t + D$ 
($D$ being a relative deadline);
The $Job$ is not initially in $run$ mode 
(but if it has high priority, this might happen almost immediately).

\item In general, a $JOB$ will go through a number of cycles out/in of $run$ mode;

\item $P\sb{e}$ forces recording in its $e$ the sum of time it is executing.

\item When a $JOB$ finishes, it is removed from $active$.

\end{itemize}

%\cnote{Check next}
Notice that starting a Job does not make it run ---
that is the role of the $Scheduler$.
Thus, jobs start and are added to $active$;
the scheduler sets and resets their $run$ field.

%%%%%%%%%%%%%%%
%\subsection{\rm\bf Static scheduling issues}
%\label{S-static-sched}
%
%\alannote{Human scheduler \ldots\\
%input\\
%guar}

%%%%%%%%%%%%%%
\subsection{Dynamic Scheduler}
\label{S-dyn-sched}
         
The $run$ field of a $Job$ records whether it is actively executing.
In fact, 
the only way the Scheduler can allow a job to progress,
is change its $run$ filed and
%\footnote{In some resilient modes,
%the scheduler can re-allocate  ``budgets''
%and/or terminate jobs.}
this makes the way the scheduler achieves its specification rather indirect
in that it relies on $P\sb{e}$.

The specifications are expressed at two levels:
an optimistic mode in which all jobs can be scheduled successfully and a resilient mode in which lo-crit jobs 
can be abandoned if necessary to meet the deadlines of hi-crit jobs.

%%%%%%%%%%%%%%
\subsubsection{Optimistic mode}

The overall scheduling objective is to make sure that jobs can finish by their respective deadlines;
if all jobs were of equal criticality,
this could be expressed by requiring that the following optimistic invariant is maintained:

\begin{fn}{inv-State\sb{O}}{mk-State(t, active)}\\
\signature{State \to \Bool}
\forall{mk-Job(e, , mk-JobInfo(d, C, )) \in \rng{active}}{C \minus e  \leq d \minus t}
\end{fn}

\noindent
This states  that all jobs currently in the system must have room to finish by their deadline time ($d$).
A corollary of this is that each job would terminate no later than its deadline.%
\footnote{It is shown below that a more conservative invariant must be preserved to achieve resilience.}
%\cnote{I'd like to add something about the fact that the concern about the sum,
%over all jobs, 
%of the remaining execution time fitting has been handled in the static scheduling work!}
The condition above is necessary but not in itself sufficient:
prior schedulability analysis will have ascertained that the expected job arrival pattern 
will not be such as to make the sum of remaining execution times exceed capacity
(e.g.~it will not by possible for two jobs to arrive $n$ units of duration before their deadline and both jobs 
having worst case execution time of $n$).

Preserving $inv-State\sb{O}$ can be viewed as the key obligation of the scheduler but
the only way in which it can achieve this is by allocating time to $Job$s in $active$ 
which entails setting their $run$ field to $\true$
($P\sb{e}$ then governs the increase in their $e$ field).%
\footnote{Readers familiar with Rely-Guarantee literature might wonder if such invariants
should be couched as guarantee conditions of the form:

\begin{formula}
inv-State\sb{O}(st) \Implies inv-State\sb{O}(st')
\end{formula}

\noindent
The reason for using an invariant is that $Time$ advances even when the scheduler is inactive.
A guarantee condition puts constraints on what happens over steps of the relevant process;
maintaining the invariant imposes a stronger requirement that the scheduler does something often enough.}

To facilitate maintenance of this invariant, the rely condition of the scheduler must cover both
reliance on all $Job$s respecting their $C$ and
the $JobInfo$ values of each job are unchanging 
(remember that jobs can arrive in --and leave from-- $active$).

\begin{op}[SCHEDULER\sb{O}]
\ext{\Rd t: ClockValue\\
       \Rd active: \mapof{I}{Job}}
\rely{(\forall{mk-Job(e', , info') \in \rng{active'}}{
        e' \leq info'.C}) \And\\
        \forall*{i \in (\dom{active} \inter \dom{active'})}
              {active'(i).run = active(i).run \And
               active'(i).info = active(i).info}
        }
\end{op}

Correspondingly,
the specification of each job is defined as follows:

\begin{op}[JOB(i)]
\ext{\Rd job: Job}
\rely{job'.info = job.info}
\guar{job'.e' \leq (job.info).C}
\end{op}

\noindent
The index $i$ locates a $Job$ in the $active$ map of $State$.
 
%\cnote{Check of types:
%to follow!}

The standard Rely/Guarantee proof obligation for parallel composition requires that the rely conditions 
of each component follow from the guarantee conditions of their sibling processes.

\begin{theorem}
For $JOB$:

\begin{formula}
guar-SCHEDULER\sb{o}(st, st') \Rightarrow rely-JOB(i)((st.active)(i), st'.active)(i)) 
\end{formula}

Is immediate from their respective frames.
\end{theorem}

\begin{theorem}
For $SCHEDULER$:

\begin{formula}
\And\sb{i} guar-JOB(i)((st.active)(i), st'.active)(i)) \Rightarrow rely-SCHEDULER\sb{o}(st, st')
\end{formula}

Is also immediate.
\end{theorem}

Although the guarantee conditions of jobs imply the rely condition of the scheduler,
it cannot be implemented unless the developer can also assume that $P\sb{e}$ is satisfied.

Notice that the value $C$ has two roles: each job relies on its environment behaving according
to whatever model or measuring process was used to derive $C$, but the job also has a contract
with the scheduler not to execute for more than $C$. The scheduler is assumed to have used some form of
analysis to verify (offline usually) that if all jobs respect their guarantee conditions then it will indeed
provide the necessary capacity to each job. Hence the job can rely upon receiving $C$ before its deadline.

It should again be  noted  that this specification of the Scheduler's behaviour does not include the actual details of the scheduling algorithm
or dispatching policy --- it is just a specification of what the Scheduler must achieve (its obligations). For a specific set of jobs it may not
be possible to derive a valid scheduler that can meet this specification.

%%%%%%%%%%%%%%
\subsubsection{Resilient mode}

%\cnote{ADD:\\
Consider what happens if a job overruns its estimated $C$:
the guarantee condition of that job no longer holds and this invalidates the rely condition of the Scheduler.
If the specifications given above were the only requirement on an implementation,
the invalidated $rely-SCHEDULER$ removes the obligation of the developed Scheduler to function according to its specification.

Providing resilience requires that a weaker specification should be met in such circumstances.
There are then two issues: 
defining the weaker (less optimal) set of conditions and
ensuring a smooth transition.

In resilient mode, the scheduler only guarantees that hi-crit jobs get serviced and
lo-crit jobs might be terminated or fail to complete by their deadlines.%
%\footnote{A third possibility is that some lo-crit jobs do complete by their deadlines.
%It is just that the scheduler is not required to achieve this.}
\footnote{Paradoxically,
it would illustrate the use of nested conditions rather better if there were more criticality levels because 
each failure could cascade to a lower level.
The reason for limiting to two levels in this paper is just notational brevity.}    The specification does not require such terminations,
it simply specifies a lower bound on the Scheduler.
As indicated at the beginning of Section~\ref{S-dyn-sched},
the optimistic invariant preserves enough resource to be able to complete any job whether hi or lo-crit.
In contrast,
$inv-State\sb{R}$ enshrines a cautious approach of making sure that 
hi-crit jobs can meet their deadlines even if they all use their extra time allocation:%

\begin{fn}{inv-State\sb{R}}{mk-State(t, active, )}\\
\signature{State \to \Bool}
\forall*{mk-Job(e, , mk-JobInfo(d, C, crit)) \in \rng{active}}{
	crit \in Hi \Implies C + crit.X \minus e  \leq d \minus t}
\end{fn}
%
%\cnote{I need to change the order here and look at the failing implication before the test case!\\
%It's also worth saying that one could maintain enough ``fat'' to cover the sum of the $X$s}

As in optimistic mode,
maintaining $inv-State\sb{R}$ is a key issue for the Scheduler.
Furthermore,
the implementation is required to meet both the optimistic behaviour and provide the specified resilient mode.
This raises the issue of the transition between modes
(i.e.~what happens when the Scheduler has to change modes because of an overrunning job).

If it were the case that $inv-State\sb{O}$ implied $inv-State\sb{R}$,
things would be simple.
In that $inv-State\sb{R}$ is concerned with only hi-crit jobs,
it is a relaxation.
But these hi-crit jobs can now demand more resource so the unguarded implication does not hold.
It is necessary to find conditions under which a smooth transition can be achieved.

\begin{theorem}
Under suitable conditions:

\begin{formula}
conds(st) \And inv-State\sb{O}(st)  \Rightarrow inv-State\sb{R}(st)
\end{formula}

\noindent 
One case where it holds trivially is if all of the extra ($X$) allowances are zero.
\end{theorem}

\begin{theorem}
Another option is for the Scheduler to maintain that degree of reserve for all hi-crit jobs since:

\begin{formula}
inv-State\sb{R}(st) \And  inv-State\sb{O}(st) \Rightarrow inv-State\sb{R}(st)
\end{formula}

\noindent 
is obviously immediate.

\end{theorem}

In resilient mode ($SCHEDULER\sb{R}$), 
the weaker rely condition concerns only hi-crit jobs 
and accepts that their execution might need the extra ($X$) execution time. 
So the invariant only requires that the scheduler concerns itself with hi-crit jobs.%
%\footnote{Many schedulers live more dangerously and hold only a global reserve (``fat'') or work with $\sfrac{3}{4}$ of $X$ ---
%this would require a different specification.}

%\cnote{Need words from Alan here}

What actual Schedulers do is,
in effect, 
to retain enough slack to be able to manage transitions that are judged 
in the scheduling analysis to be important.

Consider the following example:

To see that $inv-State\sb{O}$ is not strong enough as an invariant to handle mixed criticality,
consider the following example which shows that there is an issue when resilience to overrunning hi-crit jobs is included.
Suppose there was a hi-crit job ($a$) with:
$e\sb{a} = 10$,
$d\sb{a} = 56$,
$C\sb{a} = 15$ and
$X\sb{a} = 3$.
There might also be some other lo-crit jobs so $active$ would contain:

\begin{formula}
\left\{\begin{array}{l}
a \mapsto mk-Job(10, \true, mk-JobInfo(56, 15, mk-Hi(3))),\\
b \mapsto \cdots
\end{array}
\right\}
\end{formula}

\noindent
If this situation existed at $t = 52$,
there would be insufficient time to complete the hi-crit job ($a$) by its deadline.
(Presumably some other job with an earlier deadline would have been using the resource.) 
%The lo-crit job ($b$) could also be a problem but this can anyway be abandoned in resilient mode.
A scheduler could abandon execution of any lo-crit jobs such as $b$ and employ the e$X$tra allowance stored for $a$ ---
in the example there is an extra allowance of 3 units of time to be allocated to $a$ but it is too late to meet the deadline (56)
so this does not help.
Therefore the situation must not be allowed to arise at $t = 52$ ---
it is clear that the scheduler has not kept enough ``fat'' to be able to complete $a$ by its deadline.

Notice that this represents what the scheduler must do ---
it is at liberty to attempt to schedule more jobs than required including some that are marked as lo-crit.

The specifications are:

\begin{op}[SCHEDULER\sb{R}]
\ext{\Rd t: ClockValue\\
       \Rd active: \mapof{I}{Job}}
\rely{(\forall*{mk-Job(e', , info') \in \rng{active'}}{
        info'.crit \in Hi \Implies e' \leq C + (info'.crit).X) \And}\\
        \forall*{i \in (\dom{active} \inter \dom{active'})}
              {active'(i).run \geq active(i).run \And
              active'(i).info = active(i).info}
        }
\end{op}

In resilient mode each hi-crit job promises to stay within its extended execution time; 
lo-crit jobs make no promises.

\begin{op}[JOB(i)]
\ext{\Rd e: Duration\\
       \Rd info: JobInfo}
\rely{info' \geq info}
\guar{info.crit \in Hi \Implies e' \leq C + (info.crit).X}
\end{op}

%\todo{impl must satisfy O/R}
It is important to realise that the design of the scheduler must create code that satisfies both
$SCHEDULER\sb{O}$ and $SCHEDULER\sb{R}$:
all jobs will meet their deadlines providing none are greedy and 
hi-crit jobs will get the extra resources to meet their deadlines ---
if necessary at the expense of lo-crit jobs being abandoned.

%%%%%%%%%%%%%%%%%%%%%%%%%%%%
\section{Conclusions}\label{S-con}

The formal methods basis for this paper is the timebands framework
and the rely-guarantee approach.
These ideas have previously been shown to be applicable to the specification of Cyber Physical Systems (CPS).
Time bands have been used to avoid the confusion that arises when coarse-grained concepts are discussed at a fine granularity.
Rely-guarantee conditions both help separate documentation of components and
--when nested--
help distinguish resilient modes of operation from the ideal behaviour of a system.

The targeted application of this paper is Mixed Criticality Scheduling;
tackling MCS has required an important extension to the previously used notations.

%%%%%%%%%%%%%%
\subsection{Summary}

%\cnote{Contribution = sep of $Time$}
The objective in writing this paper was to take existing ideas on the timeband framework and rely-guarantee approaches
and to extend them so that they can be useful in specifying MCS.
To achieve this,
a novel approach to viewing time as a separate index has been proposed:
assumptions about the relationship between actual $Time$ and what happens inside a computer are recorded
essentially as guarantee conditions on the unstoppable progress of $Time$.

%\cnote{ more on nested R/G}
With MCS,
the trigger that indicates that ideal behaviour cannot be achieved is related to time:
if hi-crit jobs need more resource than their optimistic estimates,
a scheduler has to take action such as abandoning lo-crit jobs.
Layered rely-guarantee conditions cope well with describing such nested modes.

The important task of relating the passage of $Time$ in the real world with what is going on inside the computer
has been handled by having a model in which there is a function from $Time$ to machine states;
this function cannot be affected by programs
but programs do bring about discrete changes in the machine states.
Crucially,
the relationship between the $Time$ index and values in the machine states is defined by a predicate 
that can be thought of as a guarantee condition.
Appropriate definition of the precision of the time band concerns issues such as clock accuracy and drift.

%%%%%%%%%%%%%%
\subsection{Further work}
\label{S-impl}

%%%%%%%%%%%%%%
%\subsection{Some possible extensions}
%\label{S-extns}

Of course,
much work remains to be done.
%\alannote{Tasks - to be handled in a separate paper\\
This introductory paper is restricted to scheduling jobs. 
Sketches for describing the ``task level'' using the same formal ideas exist and show that they appear to suffice.
This material will be the subject of a companion paper
%\alannote{mode changes/recovery in a parallel paper}
that will also say more on the transition between modes.

There is extensive published work on MC scheduling and implementation, 
but little on their formal specification. 
The Rely-Guarantee (R/G) approach has proved to be a useful formalism for specifying non-real-time
safety-critical systems and the main contribution here is to extend R/G to (i) time and (ii) multiple criticalities. 
Such a formalism will prove to be essential since the notion of mixed criticality has subtle semantics: 
concepts such as correctness, resilience and robustness are rarely straightforward or intuitive for such systems.

It is important to remember that the material in the body of this paper concerns only specifications.
These specifications set a necessary condition on implementations but they are at liberty to achieve more.
%\todo{I think that I can see how to model Alan's 98/100\\
%- Basically: there would have to be a state component that (by a rely condition) counts the number of Jobs that are beyond their basic allowance\\
%- then the trigger for switching to hi-crit mode can be made on the cardinality of new state component
%}
A typical scheduler will ensure that jobs with the earliest deadline are executed first (EDF);
more useful schedulers might abandon lo-crit jobs in stages if so doing provides enough resource for the hi-crit jobs.
%or might say that only some (high) proportion of the hi-crit jobs meet their deadlines.
The usefulness of the specifications in the body of this paper has to be judged by
seeing how easy it is to show that such implementations satisfy the specifications proposed here.

%Probabilities}
An ambitious extension that has not yet been worked on would be to specify probability distributions on assumptions about timing and on the commitment to timely job completion.

\subsection*{Acknowledgements}

This paper relates to one aspect of research that is on going with Sanjoy Baruah and Iain Bate.
The authors also acknowledge useful suggestions made by Ian Hayes on an earlier draft of this paper.

This research has been supported in part by EPSRC (UK) grants, STRATA and MCCps and Leverhulme grant RPG-2019-020.

\bibliographystyle{alpha}

\bibliography{master,C-extras}

\end{document}